\documentclass[showpacs,preprintnumbers,prd,showkeys,aps]{revtex4}

\usepackage{braket}
\usepackage{eurosym}
\usepackage{amssymb}
\usepackage{amsfonts}
\usepackage{amsmath}
\usepackage{graphicx}
\usepackage{calrsfs}
\usepackage[colorlinks=true,
            linkcolor=red,
            urlcolor=red,
            citecolor=blue]{hyperref}

\begin{document}

\title{Dirac Particles Tunneling from Black Holes with Topological Defects}
\author{Kimet Jusufi}
\email{kimet.jusufi@unite.edu.mk}
\affiliation{Physics Department, State University of Tetovo, Ilinden Street nn, 1200 Tetovo,
Macedonia}
\date{\today }

\begin{abstract}
We study Hawking radiation of Dirac particles with spin--$1/2$ as a tunneling process from Schwarzschild--de Sitter and Reissner--Nordstr\"{o}m--de Sitter black holes in background spacetimes with a spinning cosmic string and a global monopole. Solving Dirac's equation by employing the Hamilton--Jacobi method and WKB approximation we find the corresponding tunneling probabilities and the Hawking temperature. Furthermore, we show that the Hawking temperature of those black holes remains unchanged in presence of topological defects in both cases.
\end{abstract}
\maketitle

\section{Introduction}
After the original Hawking derivation of black holes temperature \cite{hawking}, a number of different approaches were introduced, among others, the Wick rotation method \cite{gibbons,gibbons1}, quantum tunneling \cite{perkih}, anomaly method \cite{iso}, and the technique of dimensional reduction \cite{umetso}. In the tunneling method, Hawking radiation can be viewed as a tunneling process by using the semi--classical WKB approximation. In this way, the particle can quantum mechanically tunnel through the horizon and consequently can be observed at infinity as a real particle. The tunneling rate is related to the imaginary part of the action in the classically forbidden region. Generally, there are two methods to obtain the imaginary part of the action. In the first method used by Parikh and Wilczek, the imaginary part of the action is calculated by integrating the radial momentum of the particles. In the second method \cite{angheben,sri}, the imaginary part of the action is obtained by solving the relativistic Hamilton--Jacobi equation. The quantum tunneling method has been studied in great details for a number of spherically symmetric and stationary spacetimes black holes and also for different particles, including, a scalar particles, Dirac particles, spin--$1$ particles, and spin--$3/2$ particles \cite{Mann3,kimet1,sakalli1,xiang,Kruglov,sakalli2}. In particular, the tunneling from the rotating Kerr black hole, Kerr--Newman black hole \cite{Mann1,Mann2,zhang,jiang}, Kerr de Sitter and Kerr--Newman de Sitter black hole \cite{chen}, black strings \cite{ahmed,kimet}, black holes with NUT parameter \cite{gohar} and many others.

In Ref. \cite{aryal}, the metric for a static cosmic string passing throught a Schwarzschild black hole is given. Furthermore, it is shown that Hawking temperature remains unchanged due to the presence of a cosmic string. In Ref. \cite{ren}, the tunneling of scalar particles from Schwarzschild black hole with a static cosmic string and a global monopole were treated separately using the Parikh--Wilczek method. In Ref. \cite{kimet1} the tunneling of scalar particles from Schwarzschild-de Sitter (SdS) and Reissner--Nordstr\"{o}m--de Sitter (RNdS) is investigated using Hamilton--Jacobi method. The aim of this paper is to extend and generalize these results and analyze the effect of a rotating cosmic string plus a global monopole on the Hawking temperature of SdS and RNdS black holes. In particular, we invastigate the tunneling of Dirac particles with spin--$1/2$ in the presence of a cosmological constant using the Hamilton--Jacobi method. Therefore, we first consider the tunneling from the SdS and then we generalize our results for RNdS black holes pierced by an infinitely long spinning cosmic string plus a global monopole.  Topological defects may have been produced by the phase transition in the early universe. A spinning cosmic string is characterized by the rotational parameter $"a"$ and the angular momentum parameter $J$, given by $a =4J$. The spacetime of cosmic string gives rise to a number of interesting phenomena, cosmic string can act as a gravitational lens \cite{gott}, it can induce a finite electrostatic self-force on an electric charged particle \cite{linet}, shifts in the energy levels of a hydrogen atom \cite{bezerra}, they were also suggested as an explanation of the anisotropy of the cosmic microwave background radiation \cite{kaiser}. Morover, in Ref. \cite{hawking1} a possible scenario of a black hole formation from cosmic string loops was investigated.

This paper is organised as follows. In Section 2, we introduce the stationary line element near the horizon for SdS black hole in the  cosmic string and global monopole background. Then in Section 3, we calculate the tunneling rate of massive/massless Dirac particles and the Hawking temperature for the  SdS black hole. In Section 4, similary, we calculate the tunneling rate and the Hawking temperature for RNdS black hole in spacetime background with a cosmic string and a global monopole. In Section 5, we comment on our results.

\section{Tunneling from SdS black holes}
The simplest Lagrangian density which describes a global monopole is given by \cite{birrola}
\begin{equation}
\mathcal{L}=\frac{1}{2}g^{\mu \nu}\partial_{\mu}\phi^{a}\partial_{\nu}\phi_{a}-\frac{1}{4}\lambda (\phi^{a}\phi_{a}-\eta^{2})^{2},\label{1}
\end{equation}
where $\lambda$ is the self-interaction term and $\eta$ is the scale of a gauge-symmetry breaking. For a typical grand unification scale $\eta\sim 10^{16} $ GeV, which implies that $8\pi\eta^{2} \ll 1$. On the other hand $\phi^{a}$ is a triplet of scalar fields which transform under the group $O(3)$, whose symmetry is spontaneously broken to $U(1)$ given by
\begin{equation}
\phi^{a}=\eta h(r)\frac{x^{a}}{r}\label{2}
\end{equation}
with $x^{a}x^{a}=r^{2}$. Solving the Einstein field equations in spacetime with a global monopole and neglecting the mass of the monopole core, we can write the line element of Schwarzschild--de Sitter black hole with a global monople as follows \cite{gao}
\begin{equation}
\mathrm{d}s^{2}=-\left(1-\frac{2M}{r}-\frac{r^{2}}{l^{2}}\right)\mathrm{d}t^{2}+\left(1-\frac{2M}{r}-\frac{r^{2}}{l^{2}}\right)^{-1}\mathrm{d}r^{2}+r^{2}p^{2}\left(\mathrm{d}\theta^{2}+\sin^{2}\theta\, \mathrm{d}\phi^{2}\right).\label{3}
\end{equation}

The presence of a global monopole is encoded via $p^{2}=1-8\pi\eta^{2}$, $M$ is the black hole mass and $l^{2}=3/\Lambda^{2}$. Next, one can introduce also a static cosmic string in the metric \eqref{3}, which can be done by using $\mathrm{d}\phi\to b \,\mathrm{d}\phi$ \cite{aryal}, where  $b=(1-4\mu)$. Here $\mu$ is the linear mass density of the cosmic string, for grand--unified strings $\mu\sim 10^{-6}$. Since a cosmic string possesses a positive energy density $\mu > 0$, it follows that $b<1$. 

Therefore, we aim to introduce a rotating cosmic string by using a further coordinate transformation $\mathrm{d}t\to\mathrm{d}t+a\,\mathrm{d}\phi$ \cite{mazur,vilenkin}. In this way the metric \eqref{3} takes the form 
\begin{eqnarray}
\mathrm{d}s^{2}=-\left(1-\frac{2M}{r}-\frac{r^{2}}{l^{2}}\right)(\mathrm{d}t+a\,\mathrm{d}\phi)^{2}+\left(1-\frac{2M}{r}-\frac{r^{2}}{l^{2}}\right)^{-1}\mathrm{d}r^{2}+ r^{2}p^{2}\left(\mathrm{d}\theta^{2}+b^{2}\sin^{2}\theta\, \mathrm{d}\phi^{2}\right).
\label{4}
\end{eqnarray}

The above metric describes a rotating and noninteracting infinitely long cosmic string and a global monopole placed close to each other in Schwarzschild--de Sitter black hole spacetime. The presence of a cosmic string is encoded via $b=\left(1-4\mu\right)$, aligned through $\theta=0$ and $\theta=\pi/2$, parallel to $z$--axes, passing through the Schwarzschild--de Sitter black hole spacetime. Setting $M=\Lambda=\eta=0$, the spinning cosmic string metric in Minkowski spacetime is recovered \cite{mazur}
\begin{equation}
\mathrm{d}s^{2}=-\left(\mathrm{d}t+a\mathrm{d}\phi\right)^{2}+\mathrm{d}r^{2}+r^{2}\mathrm{d}\theta^{2}+r^{2}b^{2}\mathrm{d}\phi^{2}.
\end{equation}

In this paper, we will consider an idealized cosmic string with a parameter "$a$" constant with time, related to the angular momentum parameter $J$, with $a=4J$.  Solving $r^{3}+2Ml^{2}-rl^{2}=0$, one gets the black hole event horizon $r_{H}$ and the cosmological horizon $r_{C}$, given by
\begin{equation}
r_{H}=\frac{2M}{\sqrt{3\Xi}}\cos\frac{\pi+\psi}{3},\label{5}
\end{equation}
\begin{equation}
r_{C}=\frac{2M}{\sqrt{3\Xi}}\cos\frac{\pi-\psi}{3},
\end{equation}
where
\begin{equation}
\psi=\cos^{-1}(3\sqrt{3\Xi}).
\end{equation}

Here $\Xi=M^{2}/l^{2}$ and belongs to the interval $0<\Xi<1/27$. Expanding $r_ {H} $ in terms of $M$ with $\Xi<1/27$, leads to (see, e.g., \cite{rahman1})
\begin{equation}
r_{H}=2M\left(1+\frac{4M^{2}}{l^{2}}+\cdots\right),\label{6}
\end{equation}
clearly, in the limit $\Xi\to 0$, it follows  $r_{H}\to 2M$. For the sake of convenience, let us write the metric \eqref{4} near the event horizon. For that purpose, one can define $\Delta=r^{2}-2Mr-r^{4}/l^{2}$, so the line element near the event horizon becomes 
\begin{eqnarray}
\mathrm{d}s^{2}=&-&\frac{\Delta_{,r}(r_{H})(r-r_{H})}{r^{2}_{H}}(\mathrm{d}t+a\,\mathrm{d}\phi)^{2}+\frac{r^{2}_{H}}{\Delta_{,r}(r_{H})(r-r_{H})}\mathrm{d}r^{2}+r_{H}^{2}p^{2}\left(\mathrm{d}\theta^{2}+b^{2}\sin^{2}\theta \,\mathrm{d}\phi^{2}\right),\label{7}
\end{eqnarray}
where 
\begin{equation}
\Delta_{,r}(r_{H})=\frac{\mathrm{d}\Delta}{\mathrm{d}r}\bigg|_{r=r_{H}}=2\left(r_{H}-M-2\frac{r_{H}^{3}}{l^{2}}\right).\label{8}
\end{equation}

Due to the frame-dragging effect of the coordinate system in the stationary rotating spacetime, we can perform the dragging coordinate transformation $\varphi=\phi-\Omega t$, where
\begin{equation}
\Omega_{b,p}(r)=\frac{a\,\Delta_{,r}(r_{H})(r-r_{H})}{r_{H}^{4}p^{2}b^{2}\sin^{2}\theta-a^{2}\Delta_{,r}(r_{H})(r-r_{H})}.
\label{9}
\end{equation}

In this way the metric \eqref{4} can be written in a more compact form
\begin{equation}
\mathrm{d}s^{2}=-F(r)\mathrm{d}t^{2}+\frac{1}{G(r)}\mathrm{d}r^{2}+K^{2}(r)\mathrm{d}\theta^{2}+H^{2}(r)\mathrm{d}\varphi^{2},\label{10}
\end{equation}
where
\begin{eqnarray}
F(r)&=&\frac{b^{2}p^{2}r_{H}^{2}\sin^{2}\theta \Delta_{,r}(r_{H})(r-r_{H})}{b^{2}p^{2}r_{H}^{4}\sin^{2}\theta-a^{2}\Delta_{,r}(r_{H})(r-r_{H})}, \\
 G(r)&=&\frac{\Delta_{,r}(r_{H})(r-r_{H})}{r^{2}_{H}},    \\
K^{2}(r)&=&p^{2}r_{H}^{2},\\
H^{2}(r)&=&p^{2}b^{2}r_{H}^{2}\sin^{2}\theta-a^{2}\frac{\Delta_{,r}(r_{H})(r-r_{H})}{r_{H}^{2}}.
\end{eqnarray}

In what follows, we will use the metric \eqref{10}, to study the tunneling rate of spin--$1/2$ particles from the event horizon. The tunneling rate is related to the imaginary part of the action in the classically forbidden region given by \cite{sri}
\begin{equation}
\Gamma\sim\exp{\left(-2\,\mbox{Im}\,S\right)}=\exp\left(-\frac{E_{net}}{T}\right).
\end{equation}
where $S$ is the action of the classically forbidden trajectory of the tunneling particle, which
has a net energy $E_{net}$ and temperature $T$. One way for finding $S$ is to use the Hamilton--Jacobi method by choosing a suitable ansatz using the Killing vectors of
the background spacetime (spacetime symmetries). Then, one can use the relativistic Hamilton--Jacobi equation and solve for the radial integral located at the event horizon. 

\section{Tunnelig of Dirac particles from SdS black hole}
The motion of Dirac particles with mass $m$ in curved spacetime for the spinor field $\Psi$, is given by Dirac's equation \cite{mayeul}
\begin{equation}
i \gamma ^{\mu }\left(D_{\mu}\right)\Psi +\frac{m}{\hbar }\Psi =0.
\label{18}
\end{equation}

The covariant derivative is given by $$ D_{\mu}=\partial_{\mu }+\Omega_{\mu},\,\,\,\, \Omega _{\mu}=\frac{i}{2} \,\Gamma _{\mu }^{\alpha \beta }\,\Sigma _{\alpha
\beta }$$
where
\begin{eqnarray}\notag
\Gamma _{\mu }^{\alpha \beta }&=& g^{\beta \gamma}\Gamma _{\mu \gamma}^{\alpha},\\\notag
\Sigma_{\alpha \beta }&=&\frac{i}{4} \left[ \gamma
^{\alpha },\gamma ^{\beta }\right],\\
\left\lbrace\gamma^{\alpha},\gamma^{\beta}\right\rbrace&=&2g^{\alpha \beta}\times  \mathbb{I}_{(4\times 4)}.
\end{eqnarray} 

We are free to choose the $\gamma ^{\mu }$ matrices in different ways, let us for simplicity choose  $\gamma ^{\mu }$ matrices as \cite{Mann1}

\begin{eqnarray}\nonumber
\gamma ^{t}=\frac{1}{\sqrt{F(r)}}\left(
\begin{array}{cc}
i & 0 \\
0 & -i
\end{array}
\right), \,\,\,\, \gamma ^{r}=\sqrt{G(r)}\left(
\begin{array}{cc}
0 & \sigma ^{3} \\
\sigma ^{3} & 0
\end{array}
\right),             
\end{eqnarray}
\begin{eqnarray}\nonumber
\gamma ^{\theta}=\frac{1}{K(r)}\left(
\begin{array}{cc}
0 & \sigma ^{1} \\
\sigma ^{1} & 0
\end{array}
\right), \,\,\,\, \gamma ^{\varphi}=\frac{1}{H(r)}\left(
\begin{array}{cc}
0 & \sigma ^{2} \\
\sigma ^{2} & 0
\end{array}
 \right) ,            
\end{eqnarray}
where $\sigma ^{i } \,(i=1,2,3)$ are the Pauli matrices
\[
\sigma ^{1}=\left(
\begin{array}{cc}
0 & 1 \\
1 & 0
\end{array}
\right), \,\, \sigma ^{2}=\left(
\begin{array}{cc}
0 & -i \\
i &  0
\end{array}
\right), \,\,\sigma ^{3}=\left(
\begin{array}{cc}
1 & 0 \\
0 & -1
\end{array}
\right) .
\]

Therefore, we need to solve the following Dirac equation,  written as 
\begin{equation}
i\gamma ^{t}\partial _{t}\Psi+i\gamma ^{r }\partial _{r }\Psi+i\gamma ^{\theta }\partial _{\theta}\Psi+i\gamma ^{\varphi}\partial _{\varphi }\Psi +i \gamma ^{\mu }\Omega_{\mu}\Psi+\frac{m}{\hbar }\Psi =0.
\label{20}
\end{equation}

As we know, the state of Dirac particles with spin-$1/2$ is described by two corresponding states, spin-up and spin-down states. In order to solve Dirac's equation (\ref{20}) we can use the following ansatz for Dirac's field $\Psi$

\begin{equation}
\Psi _{\uparrow }\left( t,r,\theta ,\varphi\right) =\left(
\begin{array}{c}
A\left( t,r,\theta ,\varphi\right)  \\
0\\
B\left( t,r,\theta ,\varphi\right) \\
0
\end{array}
\right) \exp \left( \frac{i}{\hbar }S_{\uparrow }\right),
 \label{21}
\end{equation}
corresponding to spin up case $(\uparrow ) $, and 
\begin{equation}
\Psi _{\downarrow }\left( t,r,\theta ,\varphi\right) =\left(
\begin{array}{c}
0  \\
C\left( t,r,\theta ,\varphi\right)\\
0 \\
D\left( t,r,\theta ,\varphi\right)
\end{array}
\right) \exp \left( \frac{i}{\hbar }S_{\downarrow}\right),
 \label{22}
\end{equation}
for the spin down case $(\downarrow )$. Here, $S_{\uparrow} $ and $S_{\downarrow}$ donates the corresponding action of Dirac particles with spin $(\uparrow)$ and $(\downarrow)$, $A$ and $B$ are two arbitrary functions of the coordinates. Using the symmetries of the metric (\ref{10}), given by Killing vectors, we can choose, therefore, the following ansatz for the action in the spin up case
\begin{equation}
S_{\uparrow }\left( t,r,\theta ,\varphi \right) =-\left(E_{b,p}-J_{b,p}\Omega_{b,p}\right)\,t+R(r)+J_{b,p}\varphi+\Theta(\theta)+C\label{23}
\end{equation}
here $E_{b,p}$ is the energy of the emitted particles measured at infinity, and $J_{b,p}$ is the angular momentum of the particle and $C$ is a complex number. However, since a topological defects exits, the Komar's energy $E_{b,p}$ and angular momentum $J_{b,p}$ of the particles are decreased  by a factor of $p^{2}b$. Namely, the energy and the angular momentum of the particle are, $E_{b,p}=p^{2}b E$, and  $J_{b,p}=p^{2}b J$, respectively. We can now apply the WKB approximation by inserting the Eq. (\ref{21}) into Eq. (\ref{20}) and keep only the leading order of $\hbar $ then we end up with the following four equations: 
\begin{equation}
 -i\frac{A \left(\partial_{t}S_{\uparrow }\right)}{\sqrt{F\left( r\right) }}- B \sqrt{G\left(
r\right) } \left(\partial_{r}S_{\uparrow }\right) +mA=0,   \label{24}
\end{equation}
\begin{equation}
-B\left( \frac{\left(\partial_{\theta}S_{\uparrow }\right)}{K(r)}+\frac{i}{H(r) }\left(\partial_{\varphi}S_{\uparrow }\right)\right)=0, \label{25}
\end{equation}
\begin{equation}
 i\frac{B \left(\partial_{t}S_{\uparrow }\right)}{\sqrt{F\left( r\right) }}-A \sqrt{G \left(
r\right) } \left(\partial_{r}S_{\uparrow }\right) +mB=0,   \label{26}
\end{equation}
\begin{equation}
-A\left( \frac{\left(\partial_{\theta}S_{\uparrow }\right)}{K(r)}+\frac{i}{H(r) }\left(\partial_{\varphi}S_{\uparrow }\right)\right)
=0. \label{27}
\end{equation}

At first, it seems that Eqs. (\ref{25}) and (\ref{27}), suggest that there should be a contribution to the imaginary part of the action coming from $\Theta(\theta)$, however, one can show that the contribution of $\Theta(\theta)$ to the imaginary part of the action is cancelled out, since the contribution from $\Theta(\theta)$ is completely same for both the outgoing and ingoing solutions \cite{chen}. Hence, only the first and the third equation remains to be discussed. The radial part $R (r) $ of the action $S_{\uparrow}  $ can be calculated from the following equations:

\begin{equation}
 i p^{2}b\, A \left(E-J\Omega_{b,p} \right)-B \sqrt{F\left(r\right)G\left(r\right)} \,\partial_{r}R+mA\sqrt{F\left(r\right)}=0, \label{88}
\end{equation}
\begin{equation}
i  p^{2}b \,B \left(E-J\Omega_{b,p} \right)+A\sqrt{F\left(r\right)G\left(r\right)} \,\partial_{r}R-mB\sqrt{F\left(r\right)}=0. \label{29}
\end{equation}

Solving first for the massless case, $m=0$, we get two solutions $A=\pm i B$. Therefore, the radial part of the action reads
\begin{equation}
R_{\pm}(r)=\pm\int\frac{ p^{2}\,b\,\left(E-J\Omega_{b,p} \right)}{\sqrt{\zeta(r)}}\mathrm{d}r
\label{30}
\end{equation}
where $\zeta(r)=F\left(r\right)G\left(r\right)$. Note that $+/-$ correspond to the outgoing/ingoing solutions. Let us now solve the above integral, first we can expand the function $\zeta(r)$ in Taylor's series near the horizon 
\begin{equation}
\zeta(r_{H})\approx \zeta^{\prime }(r_{H})(r-r_{H}).
\end{equation}%

We can now introduce the Feynman $i\epsilon$-prescription and make use of the formula $(r-r_{H}-i\epsilon)^{-1}=\mathcal{P}[(r-r_{H})^{-1}]+i\pi\delta(r-r_{H})$, where $\mathcal{P}$ denotes the principal part \cite{vanzo}, we get 
\begin{equation}
R_{\pm}(r_{H})=\pm\frac{\pi i r_{H}^{2}p^{2}b\,(E-J\Omega_{b,p} (r_{H}))}{\Delta_{,r}(r_{H})}+\left(\text{real part}\right)\label{31}.
\end{equation}

Note the problem of a factor two in the above result due to the temporal contribution \cite{singleton,singleton1}. There is, however, a simple solution to this problem by letting the outside particle falls into the black hole with a $100\%$ chance of entering the black hole. It follows that the corresponding probability of the ingoing Dirac particle must be 
\begin{equation*}
P_{-}\simeq e^{-2\text{Im}R_{-}}=1.
\end{equation*}%

This also implies 
\begin{equation}
\text{Im}S_{-}=\text{Im}R_{-}+\text{Im}C=0, 
\end{equation}
therefore, $\text{Im}C=-\text{Im}R_{-}$. For the outgoing particle on the other hand we have 
\begin{equation}
\text{Im}S_{+}=\text{Im}R_{+}+\text{Im}C,
\end{equation}
but from the Eq. \eqref{31}, it follows
\begin{equation}
R_{+}=-R_{-},
\end{equation}
thus, the probability for the outgoing Dirac particle is given by 
\begin{equation}
P_{+}=e^{-2\text{Im}S_{+}}\simeq e^{-4\text{Im}R_{+}}.
\end{equation}

Finally, the tunneling rate of the particles tunneling from inside to outside the horizon is given by
\begin{equation}
\Gamma =\frac{P_{+}}{P_{-}}\simeq e^{(-4\text{Im}R_{+})}.\label{32}
\end{equation}

We can now express the imaginary part of this result near the event horizon using $r_{H}=2M(1+4M^{2}/l^{2}+\cdots)$, where the angular velocity vanishes i.e. $\Omega_{b,p}(r_{H})=0$, this leads to
\begin{equation}
 \mbox{Im}R_{+}(r_{H})=2\pi M p^{2}b\left(1+\frac{16M^{2}}{l^{2}}+\cdots\right)E.
\end{equation}

Taking the imagionary part of $R_{+}(r)$ near the horizon the tunneling rate becomes
\begin{equation}
\Gamma=\exp\left[-8\pi M p^{2}b\left(1+\frac{16M^{2}}{l^{2}}+\cdots\right)E\right]. \label{33}
\end{equation}

In order to find the Hawking temperature we have to compere the last equation with the Boltzmann factor $\Gamma =\exp{\left(-\beta \left(E_{b,p}\right)\right)}$, where $\beta =1/T_{H}$. The Hawking temperature at the event horizon from SdS black holes with topological defects reads
\begin{equation}
T_{H}=\frac{1}{8\pi M}\left(1-\frac{16 M^{2}}{l^{2}}\right).
\end{equation} \label{34}

From the last two equations it's clear that Hawking radiation deviates from pure thermality, as a consequence, there is a correction to the Hawking temperature of SdS black hole. However, this result shows that Hawking temperature is unchanged in the presence of topological defects. In the particular case, setting $l\to\infty $, i.e., $\Lambda=0$, the Hawking temperature reduces to Schwarzschild black hole temperature. For the massive case, $m\neq 0$, using Eqs. (\ref{24}) and (\ref{26}) we get
\begin{equation}
\left(\frac{A}{B}\right)^2=-\frac{i p^{2}b \left(E-J\Omega_{b,p} \right)\sqrt{F\left( r\right)G\left( r\right)}-m F\left( r\right)\sqrt{G\left( r\right)
}}{i p^{2}b \left(E-J\Omega_{b,p}  \right)\sqrt{F\left( r\right)G\left( r\right)}+m G\left( r\right)\sqrt{F\left( r\right) }}.
\end{equation}

However, near the horizon, $r_{H}=2M(1+4M^{2}/l^{2}+\cdots)$, we get $A^2=-B^2$, since $F(r_{H})=G(r_{H})=0$, yielding the same Hawking temperature as in the massless case. In other words, the mass $m$ of the particle plays no relevant role in the process of Hawking radiation.

\section{Tunneling from RNdS black holes}

In Ref. \cite{gao}, the line element of Reissner--Nordstr\"{o}m black hole with positive $\Lambda$ in the background with a global monopole is given. Using similar arguments as in the first section, we can introduce a spinning cosmic string and therefore the metric reads
\begin{eqnarray}
\mathrm{d}s^{2}=-\left(1-\frac{2M}{r}+\frac{Q^{2}}{r^{2}}-\frac{r^{2}}{l^{2}}\right)\left(\mathrm{d}t+a\,\mathrm{d}\phi \right)^{2}+\left(1-\frac{2M}{r}+\frac{Q^{2}}{r^{2}}-\frac{r^{2}}{l^{2}}\right)^{-1}\mathrm{d}r^{2}+r^{2}p^{2}\left(\mathrm{d}\theta^{2}+b^{2}\sin^{2}\theta \,\mathrm{d}\phi^{2}\right).\label{35}
\end{eqnarray}

Solving for $r^{4}-l^{2}r^{2}+2Ml^{2}r-l^{2}Q^{2}=0$, we can get the event horizon $r_{H}$ and the cosmological horizon $r_{C}$ location. Without getting into details (see, e.g., \cite{rahman2}), after expanding $r_{H}$ in terms of $M$, $Q$ and $l$, with $\Xi<1/27$, it was found 
\begin{equation}
r_{H}=\frac{1}{\alpha}\left(1+\frac{4M^{2}}{l^{2}\alpha^{2}}+\cdots\right)\left(M+\sqrt{M^{2}-Q^{2}\alpha}\right)\label{36}
\end{equation}
where $\alpha=\sqrt{1+4Q^{2}/l^{2}}$. By following the same arguments that have been used in the last section, we can define $\tilde{\Delta}=r^{2}+Q^ {2} -2Mr -r^ {4} /l^ {2} $, in this way the metric (\ref{35}) near the horizon takes a similar form as (\ref{10}), since
\begin{equation}
\Delta_{,r}(r_{H})=\frac{\mathrm{d}\tilde{\Delta}}{\mathrm{d}r}\bigg|_{r=r_{H}}=2\left(r_{H}-M-2\frac{r_{H}^{2}}{l^{2}}\right).
\end{equation}

In what follows, we will use this result to study the Dirac equation and calculate the tunneling rate of spin--$1/2$ particles from RNdS black holes in the cosmic string and global monopole background. 

The equation which has to be solved is the charged Dirac equation for a particle with mass $m$, and charge $q$, given by
\begin{equation}
i \gamma ^{\mu }\left( \partial_{\mu }+\Omega_{\mu}-\frac{i q}{\hbar }A_{\mu
}\right) \Psi +\frac{m}{\hbar }\Psi =0  \label{37}
\end{equation}
where $A_{\mu}$ is the electromagnetic four-potential given by $A_{\mu}=(A_{t},0,0,0)$. Choosing the $\gamma ^{\mu} $ matrices as before, applying WKB approximation and divide by the exponential term and multiply by $\hbar $ we end up with four equations
\vspace{1.7 cm}
\begin{equation}
 -i\frac{ A \left(\left(\partial_{t}S_{\uparrow }\right)-qA_{t}\right)}{\sqrt{F\left( r\right) }}-B \sqrt{G\left(
r\right) } \left(\partial_{r}S_{\uparrow }\right)+mA=0,   \label{38}
\end{equation}
\begin{equation}
-B\left( \frac{\left(\partial_{\theta}S_{\uparrow }\right)}{K(r)}+\frac{i}{H(r) }\left(\partial_{\varphi}S_{\uparrow }\right)\right)
=0, 
\end{equation}
\begin{equation}
 i\frac{B (\left(\partial_{t}S_{\uparrow }\right)-qA_{t})}{\sqrt{F\left( r\right) }}-A \sqrt{G\left(
r\right) } \left(\partial_{r}S_{\uparrow }\right)+mB=0,   \label{39}
\end{equation}
\begin{equation}
-A\left( \frac{\left(\partial_{\theta}S_{\uparrow }\right)}{K(r)}+\frac{i}{H(r) }\left(\partial_{\varphi}S_{\uparrow }\right)\right)
=0, 
\end{equation}

Following the same arguments that have been used in the last section, we will focus only on the first and third equation. It should also be stressed that due to the presence of topological defects, the charge of the Dirac particles also shifts $q\to p^{2}b \,q$. For massless Dirac particles $m=0$, we can find the radial part $R (r) $ of the action $S_{\uparrow} $ using Eq. (\ref{38}) and Eq. (\ref{39}), yielding
\begin{eqnarray}
i\frac{ p^{2}bA}{\sqrt{F\left( r\right) }}\left(E-J\Omega_{b,p}+qA_{t}\right)-B \sqrt{G\left(
r\right) } \left(\partial_{r}R(r)\right)&=&0, \\ \label{40} 
 -i\frac{ p^{2}bB}{\sqrt{F\left( r\right) }}\left(E-J\Omega_{b,p}+qA_{t}\right)+A \sqrt{G\left(
r\right) } \left(\partial_{r}R(r)\right)&=&0.    \label{41}
\end{eqnarray}

Similarly, we get two solutions, $A=\pm iB$, corresponding to the outgoing/ingoing solutions. The radial part of the action reads
\begin{equation}
R_{\pm}(r)=\pm\int \frac{p^{2}b\left(E-J\Omega_{b,p}+qA_{t}\right)}{\sqrt{\Sigma(r)}}\mathrm{d}r.
\end{equation}
where $\Sigma(r)=F\left(r\right)G\left(r\right)$. Note that $+/-$ correspond to the outgoing/ingoing solutions. Let us now expand the function $\Sigma(r)$ in Taylor's series near the horizon 
\begin{equation}
\Sigma(r_{H})\approx \Sigma^{\prime }(r_{H})(r-r_{H}).
\end{equation}%

Since the dragged angular velocity vanishes at the horizon, i.e., $\Omega_{b,p}(r_{H})=0$, we can integrate around the pole at the event horizon $r=r_{H}$, yielding
\begin{equation}
R_{\pm}(r_{H})=\pm\frac{\pi i r_{H}^{2}p^{2}b(E+qA_{t})}{\Delta_{,r}(r_{H})}+\left(\text{real part}\right).
\end{equation}

Following the same arguments used in the last section, for the probability of the outgoing Dirac particles we have
\begin{equation}
P_{+}=e^{-2\text{Im}S_{+}}\simeq e^{-4\text{Im}R_{+}},
\end{equation}
with the tunneling rate given by 
\begin{equation}
\Gamma =\frac{P_{+}}{P_{-}}\simeq e^{(-4\text{Im}R_{+})}.
\end{equation}

Neglecting $M^{3}$ terms and its higher order terms near the event horizon and using the electromagnetic potential of the black hole  $A_{t}=Q/r_{H}$, the imaginary part of the last equation reads 
\begin{eqnarray}
\mbox{Im}R_{+}(r_{H})& = & \frac{\pi}{2\alpha}\frac{\left(M+\sqrt{M^{2}-Q^{2}\alpha}\right)^{2}p^{2}b\left(E+qA_{t}\right)}{M(1-\alpha)+\sqrt{M^{2}-Q^{2}\alpha}}.
\label{42}
\end{eqnarray}

We therefore conclude that the tunnelling rate of the charged Dirac particles at the event horizon is
\begin{eqnarray}
\Gamma=\exp \Bigg\{-\frac{2\pi}{\alpha}\frac{\left(M+\sqrt{M^{2}-Q^{2}\alpha}\right)^{2}p^{2}b\,E}{M\left(1-\alpha\right)+\sqrt{M^{2}-Q^{2}\alpha}}\left[1-\frac{\alpha\, e \,Q}{M+\sqrt{M^{2}-Q^{2}\alpha}}\left(1-\frac{4M^{2}}{l^{2}\alpha^{2}}+...\right)\right]\Bigg\}\label{43}.
\end{eqnarray}

The Hawking temperature of Dirac particles for Reissner-Nordstr\"{o}m black holes in spacetime of topological defects can be found by comparing the last equation with the Boltzmann factor $\Gamma =\exp(-\beta E_{net})$, where $E_{net}=(E_{b,p}+q_{b,p}A_{t})$. Then it follows
\begin{eqnarray}
T_{H}=\frac{\alpha}{2\pi}\frac{M(1-\alpha)+\sqrt{M^{2}-Q^{2}\alpha}}{\left(M+\sqrt{M^{2}-Q^{2}\alpha}\right)^{2}}\label{44}
\end{eqnarray}

As expected, there are corrections to the Hawking radiation, but the Hawking temperature remains unaltered in presence of topological defects. In the particular case, setting $l\to\infty $, i.e., $\Lambda=0$, and $\alpha=1$, we recover the Hawking temperature of charged Dirac particles for RNdS black hole without topological defects \cite{perkih}. In the case of massive Dirac particles $m\neq 0$, we can use Eqs. (\ref{38}) and (\ref{39}) and get the following result
\begin{equation}
\left(\frac{A}{B}\right)^2=-\frac{i \,p^{2}b \left(E-J\Omega+qA_{t}\right)\sqrt{F\left( r\right)G\left( r\right)}-m F\left( r\right)\sqrt{G\left( r\right)
}}{i\, p^{2}b\left(E-J\Omega+qA_{t}\right)\sqrt{F\left( r\right)G\left( r\right)}+m G\left( r\right)\sqrt{F\left( r\right) }}.\label{45}
\end{equation}
At the event horizon $r_{H}=(1/\alpha) \left(1+4M^{2}/l^{2}\alpha^{2}+\cdots\right)(M+\sqrt{M^{2}-Q^{2}\alpha})$, we get $A^2=-B^2$, since $F({r_{H}})=G(r_{H})=0$, those, we have shown that the mass of the particles is irrelevant in this process.

\section{Conclusion}

In this paper, we have extended the quantum tunneling of Dirac particles with spin--$1/2$, from Schwarzschild--de Sitter and Reissner--Nordstr\"{o}m-de Sitter black holes in the spacetimes background with a spinning cosmic string and a global monopole. The Dirac's equation has been solved via  WKB approximation and using the Hamilton-Jacobi equation. Taking into account the change of the Komar's energy, angular momentum, and charge of the particles in spacetimes with topological defects, we have calculated the tunneling rate and the correspondin black holes Hawking temperature in both cases. As a result, it is shown that Hawking temperature remains unchanged for massive as well as for massless Dirac particles and unaffected by the presence of topological defects in both cases in agreement with \cite{ren,aryal,kimet1}. 
On the other hand, we have also shown that the Hawking temperature depends
on the presence of a cosmological constant in both cases. 

Surprisingly, however, the Hawking temperature seems to be independent of the rotational parameter $a= 4J$, and hence no special role played by a rotating cosmic string. In a more
general scenario, will be very interesting to see what happens when the rotational
parameter $a = 4J$, changes with time i.e., $a(t)$. This process of angular momentum loss
should be followed by a flux of particles emitted by a cosmic string and therefore
the metrics \eqref{4} and \eqref{35} are expected to be unstable, which on the other hand, may shed some light on various astrophysical scenarios. This is beyond the scope of this paper and we plan to deal with this problem in future works.

\section*{Acknowledgement}
The author would like to thank the editor and the anonymous reviewers for the very useful comments and suggestions which help us improve the quality of this paper.

\end{document}